\documentclass[showpacs,twocolumn,superscriptaddress,aps]{revtex4}%
\usepackage{amsmath}
\usepackage{graphicx}
\usepackage{amsfonts}
\usepackage{amssymb}%
\begin{document}
\title{Single-photon logic gates using minimal resources}
\author{Qing Lin}
\email{qlin@mail.ustc.edu.cn}
\affiliation{College of Information Science and Engineering, Huaqiao University (Xiamen),
Xiamen 361021, China}
\author{Bing He}
\email{bhe98@earthlink.net }
\affiliation{ Institute for Quantum Information Science, University of Calgary, Alberta T2N 1N4, Canada}


\begin{abstract}
We present a simple architecture for deterministic quantum circuits operating on single photon qubits.
Few resources are necessary to implement two elementary gates and can be recycled for computing with large numbers of qubits.
The deterministic realization of some key multi-qubit gates, such as the Fredkin and Toffoli gate, is greatly simplified
in this approach.

\end{abstract}
\maketitle

\section{\bigskip Introduction}

Quantum computing has attracted wide attention for its factoring power and efficient
simulation of quantum dynamics. Many efforts have been made in building quantum
computers with various physical systems, and optical qubits are regarded as a prominent candidate for their robustness against decoherence.
An important theoretical breakthrough in the field was the Knill-Laflamme-Milburn (KLM)
protocol \cite{KLM}, a circuit-based approach using single photon sources, single photon detectors and linear optical elements. A two-qubit gate could be realized in an asymptotically deterministic way, as the number of photons forming an entangled state for teleportation in the protocol grows to infinity \cite{KLM, KD}. It opens up the possibility of building any quantum logic gate which can be decomposed
into two-qubit and single-qubit gates theoretically \cite{Nielsen}. The prohibitively large overhead cost of a two-qubit gate in the KLM protocol,
however, necessitates various improvements. Most progress follows in the direction of one way computation \cite{Raussendorf},
an approach imprinting circuits on a particular class of entangled states (cluster states) through measurements.
Though it is possible to create cluster states with realistic optical methods \cite{P}, the generation of such multiply entangled states is
still not efficient with available techniques, imposing a bottleneck on the practical implementation.
Beyond linear optics, a near-deterministic CNOT gate based on weak nonlinearities \cite{Nemoto} has been proposed, and it suggests a way for
deterministic quantum computation \cite{Munro}. In realistic quantum computation, however, it will still require considerable resources
to perform a gate involving more than two qubits, if one decomposes a complicated quantum circuit into the basic CNOT and single-qubit gates.

An efficient quantum computation approach demanding fewer resources is desirable.
In this work, we propose an architecture for quantum logic gates operating on qubits simply encoded as the linear combinations
of two single photon modes, e.g., $|0_{L}\rangle\equiv|H\rangle$ and $|1_{L}\rangle\equiv |V\rangle$, where $H$ and
$V$ are two polarization modes. In this architecture a quantum logic gate
can be deterministically realized with a combination of two elementary
gates. Only one ancilla photon and a few coherent states, which can be recycled after implementing one elementary gate, are necessary
to compute with a large number of qubits. Because the qubits and ancillas are in simple quantum states, the operation error of the logic gates
would be largely reduced.

\begin{figure}[ptb]
\includegraphics[width=8.7cm]{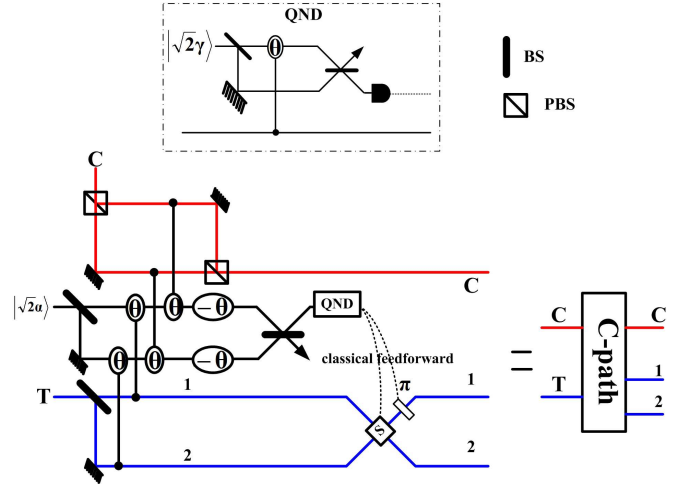}
\caption{(Color online) Schematic setup for Controlled-path gate. A PBS used in the circuit transmits the mode $|H\rangle$ and reflects the mode $|V\rangle$ of a single photon. Two qubus beams are coupled to the photonic modes as
indicated. The XPM phases on the qubus beams are $\theta$, and two phase shifter $-\theta$
 are applied to the qubus beams. The QND module in dash-dotted line is used to perform number-resolving detection.}%
\end{figure}

\section{\bigskip C-path gate}

The first ingredient in our architecture is the Controlled-path (C-path)
gate introduced in \cite{Lin}. Here, as shown in Fig. 1, we propose a design with the double
cross-phase modulation (XPM) method in \cite{He} to make it more efficient and feasible.
This gate performs the following operation on an initial two-photon state $\left\vert
\psi\right\rangle _{CT}$ ($C$ stands for the control and $T$ the target):
\begin{align}
&  \left\vert \psi\right\rangle _{CT}=a\left\vert HH\right\rangle
_{CT}+b\left\vert HV\right\rangle _{CT}+c\left\vert VH\right\rangle
_{CT}+d\left\vert VV\right\rangle _{CT}\nonumber\\
&  \rightarrow a\left\vert HH\right\rangle _{C1}+b\left\vert HV\right\rangle
_{C1}+c\left\vert VH\right\rangle _{C2}+d\left\vert VV\right\rangle
_{C2}=\left\vert \phi\right\rangle ,
\end{align}
where the index $1$ and $2$ denotes two different paths, implementing the
control on the target qubit paths by the polarizations of the control qubit.

In Fig. 1, we first use a 50:50 beam splitter (BS) to divide the target photon $T$ into two spatial modes 1 and 2.
Then two quantum bus (qubus) beams $\left\vert
\alpha\right\rangle \left\vert \alpha\right\rangle $ are introduced, with the
first coupling to the target photon mode on path 1 and the $\left\vert
V\right\rangle $ mode of the control photon, while the second to the target
mode on path 2 and the $\left\vert H\right\rangle $ mode of the control
through Kerr media, respectively. Suppose the XPM phase shifts induced in the
processes are all $\theta$. After that, a $-\theta$ phase shifter is
respectively applied to two qubus beams. Finally, one more 50:50 BS implements
the transformation $\left\vert \alpha_{1}\right\rangle \left\vert \alpha
_{2}\right\rangle \rightarrow\left\vert \frac{\alpha_{1}-\alpha_{2}}{\sqrt{2}%
}\right\rangle \left\vert \frac{\alpha_{1}+\alpha_{2}}{\sqrt{2}}\right\rangle
$, realizing the state
\begin{align}
&  \frac{1}{\sqrt{2}}\left(  \left\vert H\right\rangle _{C}\left(  a\left\vert
H\right\rangle _{2}+b\left\vert V\right\rangle _{2}\right)  \left\vert
-\beta\right\rangle +\left\vert V\right\rangle _{C}\left(  c\left\vert
H\right\rangle _{1}+d\left\vert V\right\rangle _{1}\right)  \left\vert
\beta\right\rangle \right) \nonumber\\
&  \times\left\vert \sqrt{2}\alpha\cos\theta\right\rangle +\frac{1}{\sqrt{2}%
}\left\vert \phi\right\rangle \left\vert 0\right\rangle \left\vert \sqrt
{2}\alpha\right\rangle , \label{2}%
\end{align}
where $\left\vert \beta\right\rangle =\left\vert i\sqrt{2}\alpha\sin
\theta\right\rangle $. Then, we could use the projections $\left\vert n\right\rangle \left\langle n\right\vert $
on the first qubus beam to get the proper output.
If $n=0,$ $\left\vert \phi\right\rangle $ will be projected out;  on the other hand, if $n\neq 0$, what is realized is $e^{-in\frac{\pi}{2}}\left\vert H\right\rangle _{C}\left(  a\left\vert
H\right\rangle _{2}+b\left\vert V\right\rangle _{2}\right)  +e^{in\frac{\pi
}{2}}\left\vert V\right\rangle _{C}\left(  c\left\vert H\right\rangle
_{1}+d\left\vert V\right\rangle _{1}\right)  $, which
can be transformed to $\left\vert \phi\right\rangle $ by the application of a $\pm 1$ phase factor and the controlled switch of two paths
following the classically feed-forwarded measurement results. Since the weak nonlinearity considered here is very small ($\theta\ll 1$),
only a small portion of the initial qubus beam $|\sqrt{2}\alpha\rangle$ is consumed by detection, the
unmeasured beam in the state $|\sqrt{2}\alpha \cos\theta\rangle$ will be used in following elementary gates.

After a C-path gate, the target photon will be simultaneously in two different spatial modes depending on the polarizations of the
control photon. Therefore, an operation conditioned on the control photon's polarizations can be directly performed on the spatial modes of the target. As we will demonstrate later, this gate offers a way to realize the interaction between multiple photons indirectly.

\section{\bigskip Photon number-resolving detection}

The projectors $\left\vert
n\right\rangle \left\langle n\right\vert $ required in the C-path gate could be
well approximated by a transition edge sensor (TES), a superconducting
microbolometer that has demonstrated very high detection efficiency (95\% at
$\lambda= 1550$ nm) and high photon number resolution \cite{detector}. In
practice, it is ideal to implement such number-resolving detection with simple
devices. Here, we apply the indirect measurement method in \cite{He} for the
purpose. It is a quantum non-demolition (QND) module shown inside the
dash-dotted line in Fig. 1. The process in the QND module is as follows:
\begin{equation}
\left\vert \pm\beta\right\rangle \left\vert \gamma\right\rangle \left\vert
\gamma\right\rangle \rightarrow e^{-\left\vert \beta\right\vert ^{2}%
/2}\overset{\infty}{\underset{n=0}{{\sum}}}\frac{\left(  \pm\beta\right)
^{n}}{\sqrt{n!}}\left\vert n\right\rangle \left\vert \gamma e^{in\theta
}\right\rangle \left\vert \gamma\right\rangle .
\end{equation}
With a 50:50 BS we will obtain a set of coherent-state components $\left\vert
\frac{\gamma e^{in\theta}-\gamma}{\sqrt{2}}\right\rangle \left\vert
\frac{\gamma e^{in\theta}+\gamma}{\sqrt{2}}\right\rangle $ for $n=0,1,\cdots
,\infty$. If the amplitude $\left\vert \gamma\right\vert $ is large enough,
the photon number Poisson distributions of the states $\left\vert \frac{\gamma
e^{in\theta}-\gamma}{\sqrt{2}}\right\rangle $ will be separated with negligible overlaps.
If the dominant distribution for the component of
$n=k$ is from $n_{k}$ to $n_{k}^{^{\prime}}$, we could use a realistic
detector described by the following POVM elements to detect $\left\vert
\frac{\gamma e^{ik\theta}-\gamma}{\sqrt{2}}\right\rangle $:
\begin{align}
\Pi_{0}  &  =\overset{\infty}{\underset{n=0}{{\sum}}}\left(  1-\eta\right)
^{n}\left\vert n\right\rangle \left\langle n\right\vert ,\nonumber\\
\Pi_{n_{k}}  &  =\overset{n_{k}^{^{\prime}}}{\underset{n=n_{k}}{{\sum}}%
}\left[  1-\left(  1-\eta\right)  ^{n}\right]  \left\vert n\right\rangle
\left\langle n\right\vert ,\nonumber\\
\Pi_{E}  &  =I-\Pi_{0}-\sum_{k=1}^{\infty}\Pi_{n_{k}},
\label{POVM}
\end{align}
where $\eta<1$ is the quantum efficiency of the detector. $\Pi_{0}$ here
corresponds to detecting no photon, $\Pi_{E}$ to the response to the
negligible overlaps, and $\Pi_{n_{k}}$ to the reaction to the $k$-th Poisson
curve, respectively. The operators $\Pi_{n_{k}}$ therefore select out the
components $\left\vert k\right\rangle $ in $\left\vert \pm\beta\right\rangle $
indirectly. Physically, $\Pi_{n_{k}}$ are the different responses (by the
induced currents or voltages) of a number-non-resolving detector to
$\left\vert \frac{\gamma e^{ik\theta}-\gamma}{\sqrt{2}}\right\rangle $ of the
different intensities.

\begin{figure}[ptb]
\includegraphics[width=7.8cm]{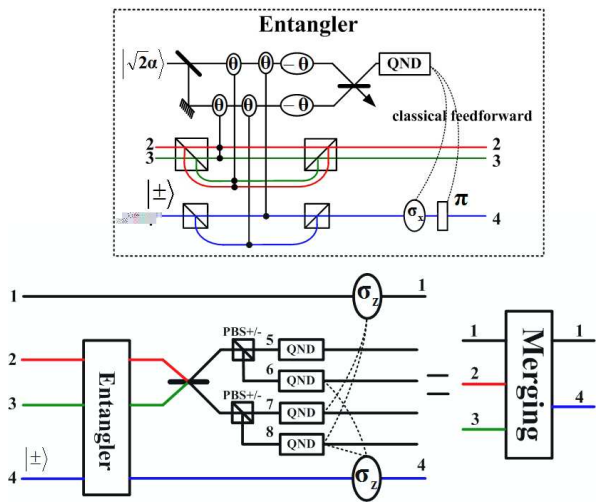}
\caption{(Color online) Schematic setup
for Merging gate. The ancilla photon in the state $|\pm\rangle$, the
single photon modes on path 2 and 3 interact with the qubus beams in the
pattern indicated in Entangler. On the first (second) qubus beam we use one XPM rotation $\theta$ to represent the coupling to $|V\rangle$
($|H\rangle$) mode on both path $2$ and $3$. A BS and four PBS$_{\pm}$ divide the modes
of the second photon to 4 paths, on which the QND modules in Fig. 1 are to
detect the photon. The operations $\sigma_z$ are performed according to the feed-forwarded measurement results of the
QNDs.}%
\end{figure}

\section{\bigskip Merging gate}

Since the C-Path operation splits the target photon into two different paths, we also
require a gate to merge these two paths back into a single path. To perform the conversion
deterministically, we introduce another elementary gate called the Merging gate shown in Fig. 2, where an
extra ancilla photon is used. It implements the transformation
\begin{align}
\left\vert \psi\right\rangle  &  =a\left\vert HH\right\rangle _{12}%
+b\left\vert HV\right\rangle _{12}+c\left\vert VH\right\rangle _{13}%
+d\left\vert VV\right\rangle _{13}\nonumber\\
&  \rightarrow a\left\vert HH\right\rangle _{14}+b\left\vert HV\right\rangle
_{14}+c\left\vert VH\right\rangle _{14}+d\left\vert VV\right\rangle _{14},
\end{align}
i.e., the merging of the second photon modes on path 2 and 3 to path 4.

The ancilla photon is in the state $\left\vert \pm\right\rangle =\frac
{1}{\sqrt{2}}\left(  \left\vert H\right\rangle \pm\left\vert V\right\rangle
\right)  $. A total state $\left\vert \psi\right\rangle \left\vert
+\right\rangle $, for example, is first sent to Entangler in Fig. 2, where we
let the photons interact with the qubus beams (see the setup in the dashed
line of Fig. 2). Similar to the double XPM pattern in C-path gate, the total
state $\left\vert \psi\right\rangle \left\vert +\right\rangle $ will be
transformed to
\begin{align}
&  \frac{1}{\sqrt{2}}a|H\rangle_{1}(|+\rangle_{2}+|-\rangle_{2})|H\rangle
_{4}+\frac{1}{\sqrt{2}}b|H\rangle_{1}(|+\rangle_{2}-|-\rangle_{2}%
)|V\rangle_{4}\nonumber\\
&  + \frac{1}{\sqrt{2}}c|V\rangle_{1}(|+\rangle_{3}+|-\rangle_{3}%
)|H\rangle_{4}+\frac{1}{\sqrt{2}}d|V\rangle_{1}(|+\rangle_{3}-|-\rangle_{3})
|V\rangle_{4}%
\end{align}
with a bit flip $\sigma_{x}$ and a phase shifter $\pi$ conditioned on the results
of the number-resolving detection on a qubus beam (no action should be taken
if $n=0$). After the interference of the modes on path 2 and 3, $|\pm
\rangle_{2}\rightarrow\frac{1}{\sqrt{2}}(|\pm\rangle_{2}+|\pm\rangle_{3})$ and
$|\pm\rangle_{3}\rightarrow\frac{1}{\sqrt{2}}(|\pm\rangle_{2}-|\pm\rangle
_{3})$, through a 50:50 BS, two PBS$_{\pm}$ let the components $\left\vert
+\right\rangle $ be transmitted while having the components $\left\vert
-\right\rangle $ reflected, making the single photon run on four different
paths numbered from $5$ to $8$. We then use the QND modules, which are the
same as that in Fig. 1, on each path to determine where the single photon in
the state $|\pm\rangle$ passes. The QND detections therefore project out the
outputs, and the projected out photon on one of the paths can be used again in
the next Merging gate.

\section{\bigskip Two-qubit gates}

Two-qubit gates such as CNOT, CZ and C-phase, which are included in the class
$|H\rangle\langle H|\otimes U_1+ |V\rangle\langle V|\otimes U_2$, can be simply constructed with these elementary gates.
Any gate operation in this form is performed by a C-path gate
followed by the single qubit operations $U_1$ and $U_2$ on the different paths of the target photon and then
a Merging gate. Compared with the qubus mediated CNOT gate in \cite{Nemoto}, a CNOT gate constructed with a pair of C-path and Merging gate uses the same amount of resources---two elementary gates and one ancilla single photon---without counting the QND modules for resolving the photon numbers in a qubus beam and preserving the ancilla photon in detections. Since we apply the double XPM method in \cite{He}, the number of the conditional XPM phase rotations in each elementary gate will be greater than that in \cite{Nemoto}. In addition to recycling the qubus beams, the advantage of the double XPM method is that a minus XPM phase shift $-\theta$, which is impractical to realize \cite{Kok}, can be avoided. Moreover, unlike the scheme in \cite{Munro}, there is no need for the displacement operations on the qubus beams, which could be hard to implement if the displacement amplitude is large \cite{Paris, loss}.

For an arbitrary two-qubit gate $U\in U\left(  4\right)$,
which is expressed as $U=\left(  A_{1}\otimes A_{2}\right)  \cdot N\left(  \alpha,\beta
,\gamma\right)  \cdot\left(  A_{3}\otimes A_{4}\right)  $, where $A_{i}\in
U\left(  2\right) $ and $N\left(  \alpha,\beta,\gamma\right)  =\exp[i\left(
\alpha\sigma_{x}\otimes\sigma_{x}+\beta\sigma_{y}\otimes\sigma_{y}%
+\gamma\sigma_{z}\otimes\sigma_{z}\right)  ]$ \cite{TG}, we can diagonalize $N\left(  \alpha,\beta,\gamma\right)$
to
\begin{align}
& |H\rangle\langle H|\otimes diag (e^{i\left(  \alpha-\beta+\gamma\right)
},e^{-i\left(  \alpha-\beta-\gamma\right)  }) \nonumber\\
&+|V\rangle\langle V|\otimes diag (e^{i\left(  \alpha+\beta
-\gamma\right)  },e^{-i\left(  \alpha+\beta+\gamma\right)})
\end{align}
with so-called magic transformation $\mathcal{M}$ \cite{TD}. The magic transformation,
which is equivalent to a CNOT and a few single qubit operations \cite{TD}, is also implementable
with C-path and Merging gates.

\section{\bigskip Multi-qubit gates}

It is straightforward to generalize to multiple
qubit gates, as any multi-qubit gate is decomposable to a product of two-qubit
gates and single-qubit gates \cite{Nielsen}. In the framework of realizing
quantum computation with the two above-discussed elementary gates, however, the
design of a multi-qubit gate can be simplified much further. We illustrate the
point with two typical multi-qubit gates---the Fredkin gate and the Toffoli gate.

The schematic setup in Fig. 3 is a Fredkin gate which implements a
swap operation on two target photons controlled by the $\left\vert
V\right\rangle $ of the control photon. In other words, it performs the
following transformation of a triple photon state $\left\vert \Psi
\right\rangle _{CT_{1}T_{2}}$:
\begin{align}
\left\vert \Psi\right\rangle _{CT_{1}T_{2}}  &  =A_{1}\left\vert
HHH\right\rangle +A_{2}\left\vert HHV\right\rangle +A_{3}\left\vert
HVH\right\rangle \nonumber\\
&  +A_{4}\left\vert HVV\right\rangle +A_{5}\left\vert VHH\right\rangle
+A_{6}\left\vert VHV\right\rangle \nonumber\\
&  +A_{7}\left\vert VVH\right\rangle +A_{8}\left\vert VVV\right\rangle
\nonumber\\
&  \rightarrow A_{6}\left\vert VVH\right\rangle +A_{7}\left\vert
VHV\right\rangle +rest.,
\label{Fred}
\end{align}
where $rest.$ denotes the unchanged terms. Here we use two C-path gates to map
the second photon to path 2 and 3, and the third photon to path 4 and 5:
\begin{align}
&  \left\vert H\right\rangle _{1}\left(  A_{1}\left\vert HH\right\rangle
+A_{2}\left\vert HV\right\rangle +A_{3}\left\vert VH\right\rangle
+A_{4}\left\vert VV\right\rangle \right)  _{24}\nonumber\\
&  +\left\vert V\right\rangle _{1}\left(  A_{5}\left\vert HH\right\rangle
+A_{6}\left\vert HV\right\rangle +A_{7}\left\vert VH\right\rangle
+A_{8}\left\vert VV\right\rangle \right)  _{35}.
\end{align}
A deterministic Fredkin gate can be therefore realized by exchanging the modes
on path 3 and 5, and using two Merging gates as the inverse operation of two
C-path gates. In the implementation of two Merging gates, only one ancilla
photon is necessary since it can be used again after QND detection. This
feature is especially useful when there are many Merging gates in computation.

\begin{figure}[ptb]
\includegraphics[width=6cm]{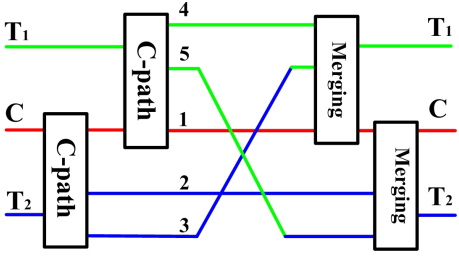}\
\caption{(Color online) Schematic setup
 for the Fredkin gate. Two C-path and Merging gates, together with the exchange of
two path modes, are used to realize a Fredkin gate directly.
 }%
\end{figure}

\begin{figure}[ptb]
\includegraphics[width=5.5cm]{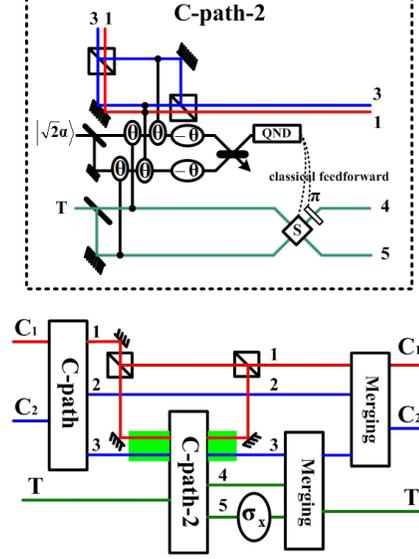}\
\caption{(Color online) Schematic setup for Toffoli gate. The H mode of photon $C_1$ and the modes on
path 3 after the first C-path gate control the target photon in the second C-path gate. A bit flip is performed on the target mode on path 5. The couplings of the qubus beams with the relevant photonic modes in C-path-2 are illustrated in dashed line. On the second qubus beam of the C-path-2 gate, we use one XPM rotation $\theta$ to represent the couplings to both $|H\rangle$ mode of path 1 and 3. }%
\end{figure}

The Toffoli gate illustrated in Fig. 4 is the bit flip of a target
photon conditioned on both $\left\vert V\right\rangle $ of two control
photons, i.e., a triple-qubit operation $(I\otimes I-|VV\rangle\langle
VV|)\otimes I+|VV\rangle\langle VV|\otimes\sigma_{x}$, where $I=|H\rangle
\langle H|+|V\rangle\langle V|$. To implement the gate, we start with a C-path gate for the initial
state $|\Psi\rangle_{C_{1}C_{2}T}$ (in the same form as $|\Psi\rangle_{C_{1}T_1T_2} $ in Eq. (\ref{Fred}))
to send the second photon $C_{2}$ to two different paths $2$ and $3$ under the control of the polarizations
of the first photon $C_{1}$:
\begin{align}
|\Psi\rangle_{C_{1}T_1T_2} &\rightarrow (A_{1}\left\vert HHH\right\rangle +A_{2}\left\vert HHV\right\rangle+A_{3}\left\vert HVH\right\rangle \nonumber\\
&+ A_{4}\left\vert HVV\right\rangle)_{12T}+(A_{5}\left\vert VHH\right\rangle +A_{6}\left\vert VHV\right\rangle \nonumber\\
&+A_{7}\left\vert VVH\right\rangle +A_{8}\left\vert VVV\right\rangle )
_{13T}=|\Psi_1\rangle. \label{Tf-1}
\end{align}
Meanwhile, as shown in dashed line of Fig. 4, a 50:50 BS divides the target photon $T$ into two paths
$4$ and $5$. Two qubus beams $\left\vert \alpha\right\rangle \left\vert
\alpha\right\rangle $ will be applied to perform the second C-path gate with
the control of the modes $|H\rangle_{3}$ and $|V\rangle_{3}$ on path $3$.
There is a slight difference in this C-path gate (denoted as C-path-2 in Fig. 4) from a standard one in Fig. 1---the second beam is coupled not only to the mode on path $5$ and the
$\left\vert H\right\rangle $ mode on path $3$ but also to $|H\rangle_{1}$ of
the first control photon, while the first beam interacts with the target mode
on path $4$ and the $\left\vert V\right\rangle $ control mode on path $3$, as indicated in the following transformation:
\begin{widetext}
\begin{align}
|\Psi_1\rangle |\alpha\rangle|\alpha\rangle &\rightarrow \frac{1}{\sqrt{2}}|HH\rangle_{12}\{ (A_1|H\rangle_4+|A_2|V\rangle_4)|\alpha e^{i\theta},\alpha e^{i\theta}\rangle+(A_1|H\rangle_5+|A_2|V\rangle_5)|\alpha,\alpha e^{i2\theta}\rangle\}
\nonumber\\
&+\frac{1}{\sqrt{2}}|HV\rangle_{12}\{ (A_3|H\rangle_4+|A_4|V\rangle_4)|\alpha e^{i\theta}, \alpha e^{i\theta}\rangle+(A_3|H\rangle_5+|A_4|V\rangle_5)|\alpha,\alpha e^{i2\theta}\rangle\}\nonumber\\
&+\frac{1}{\sqrt{2}}|VH\rangle_{13}\{ (A_5|H\rangle_4+|A_6|V\rangle_4)|\alpha e^{i\theta},\alpha e^{i\theta}\rangle+(A_5|H\rangle_5+|A_6|V\rangle_5)|\alpha,\alpha e^{i2\theta}\rangle\}\nonumber\\
&+\frac{1}{\sqrt{2}}|VV\rangle_{13}\{ (A_7|H\rangle_4+|A_8|V\rangle_4)|\alpha e^{i2\theta},\alpha\rangle+(A_7|H\rangle_5+|A_8|V\rangle_5)
|\alpha e^{i\theta},\alpha e^{i\theta}\rangle\}.
\end{align}
\end{widetext}
The remaining operations on two qubus beams are the same as those in a standard C-path gate---two phase shifts $-\theta$ and a 50:50 BS.
According to the number-resolving detection results on one qubus beam,
an output state,
\begin{align}
&  \left\vert H\right\rangle _{1}\left(  A_{1}\left\vert HH\right\rangle
+A_{2}\left\vert HV\right\rangle +A_{3}\left\vert VH\right\rangle
+A_{4}\left\vert VV\right\rangle \right)  _{24}\nonumber\\
&  +\left\vert VH\right\rangle _{13}\left(  A_{5}\left\vert H\right\rangle
+A_{6}\left\vert V\right\rangle \right)  _{4}+\left\vert VV\right\rangle
_{13}\left(  A_{7}\left\vert H\right\rangle +A_{8}\left\vert V\right\rangle
\right)  _{5},
\end{align}
will be deterministically projected out.
After that, a bit flip $\sigma_{x}$ is performed on path $5$ alone.
The total operation for a deterministic Toffoli gate will be then completed with two
Merging gates for the modes on path $4$ and $5$, and on path $2$ and $3$, respectively.

\begin{figure}[ptb]
\includegraphics[width=8cm]{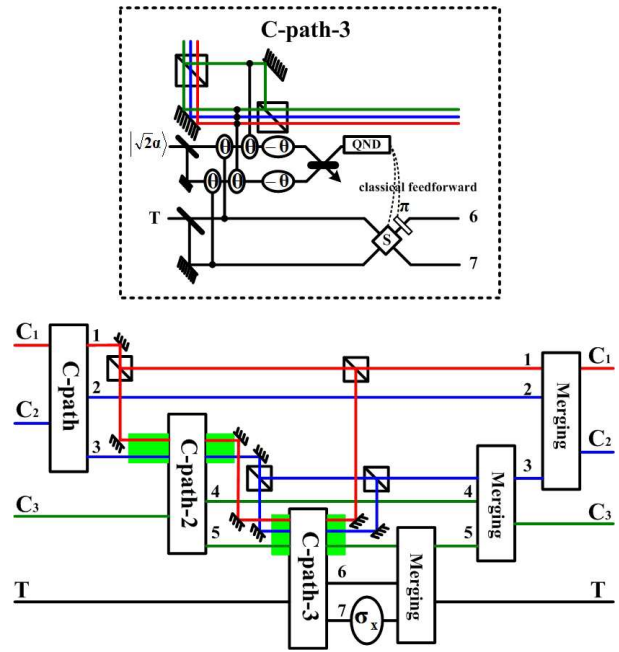}\
\caption{(Color online) Schematic setup for a triple-control Toffoli gate, with the $|V\rangle$ modes of three photons from $C_1$ to $C_3$ controlling the bit flip $\sigma_x$ on the target photon $T$. As a simple generalization of the design in Fig. 4, it applies another modified C-path gate---C-path-3---illustrated in dashed line. On the second qubus beam of the C-path-3 gate, we use just one XPM rotation $\theta$ to represent the couplings to $|H\rangle$ mode of path 1, 3 and 5, respectively. Only one ancilla photon will be necessary for the three Merging gates if we apply the detections with QND modules.}%
\end{figure}

This design can be generalized to the situation of more than three qubits, where we could simply adopt the similar coupling patterns for the photonic modes in the successive C-path gates. In Fig. 5 we outline a triple-control Toffoli gate of such type,
which implements the bit flip of a target photon under the $|V\rangle$ modes of three control photons together. The simplicity of this approach stands
out as compared with the conventional method of decomposing a quantum circuit into double-qubit and single-qubit gates.
By the conventional method, there should be at least five two-qubit gates for the Fredkin and Toffoli gates \cite{SD}.
Here we deterministically realize them with only two pairs of C-path and Merging gates, which are equivalent to two double-qubit gates.
Generally, there should be $O(n^2)$ two-qubit gates to simulate a multi-control gate of $n$ qubits in the decomposition approach \cite{B}.
As shown in Fig. 4 and 5, however, we will only need a number of the elementary gates, which grows linearly with the number of the involved qubits, to realize a multi-control gate. The decomposition of a multi-qubit circuit into two-qubit and single-qubit gates is also theoretically complicated.
But in our approach the construction of a multi-control gate follows a regular way as from Fig. 4 to Fig. 5.

\section{\bigskip Experimental feasibility}

Here we take a brief look at the feasibility of this quantum computation approach. A core technique for realizing the
elementary gates is the XPM in Kerr media. Good candidates for weak cross-Kerr
nonlinearity without self-phase modulation effects are atomic systems working
under electromagnetically induced transparency (EIT) conditions \cite{EIT}.
With light-storage technique, for example, it is possible to realize a considerable XPM
phase shift at the single photon level \cite{chen-yu}. In principle, however,
we only need a small XPM phase shift which can be compensated by the large
amplitude of the qubus beams as in \cite{Nemoto, Munro}. The error probability of a detection in the QND modules is
\begin{align}
&||\sum_{n=0}^{\infty}e^{-\left\vert \beta\right\vert ^{2}%
/2}\frac{\left(  \pm\beta\right)
^{n}}{\sqrt{n!}}\left\vert n\right\rangle(\Pi_{0}^{\frac{1}{2}}|\frac{\gamma
e^{in\theta}-\gamma}{\sqrt{2}}\rangle)||^2\nonumber\\
&\sim exp\{-2(1-e^{-\frac{1}{2}\eta\gamma^2 \theta^2})\alpha^2\sin^2\theta\},
\end{align}
rendering a near-deterministic performance given $\left\vert \alpha\right\vert
\sin\theta\gg1$. Moreover, there is no XPM phase shift $-\theta$ and no displacement operation on the qubus beams.
The design is robust against small losses of the photonic modes in the double XPM processes \cite{loss}, and is workable with the
realistic number-non-resolving detectors as we apply the indirect detection
of photon numbers.

\section{\bigskip Summary}

The architecture of a quantum computer based on two elementary gates---Controlled-path gate and Merging gate---is relatively simple compared with those of all other approaches. The data to be processed is directly encoded in single photon modes, and the ancilla photon and communication beams are also in simple quantum states so that the possibility for operational errors could be minimized. The recyclable ancillas keep the resources required in multi-qubit computing minimal. Such quantum computer may come into being with the development of the techniques of cross-Kerr nonlinearity and quantum memory for single photon qubits.

\begin{acknowledgments}
B. H. thanks C. F. Wildfeuer for helpful discussion on photon number resolving detections, and the partial support from iCORE.
\end{acknowledgments}

\end{document}